\documentclass[acus]{pac99}
\usepackage{epsfig}

\newcommand{\ud}{\mathrm{d}}
\begin{document}
\title{NONLINEAR ACCELERATOR PROBLEMS VIA WAVELETS:\\
3. EFFECTS OF INSERTION DEVICES ON BEAM DYNAMICS}
\author{A.~Fedorova, M.~Zeitlin, IPME, RAS, St.~Petersburg, Russia
\thanks{e-mail: zeitlin@math.ipme.ru}
\thanks{http://www.ipme.ru/zeitlin.html;
        http://www.ipme.nw.ru/zeitlin.html}}
\maketitle
\begin{abstract}
In this series of eight papers  we
present the applications of methods from
wavelet analysis to polynomial approximations for
a number of accelerator physics problems.
In this part, assuming a sinusoidal field variation, we consider
the analytical treatment of the effects of insertion devices on beam
dynamics. We investigate via wavelet approach a dynamical model
which has polynomial nonlinearities and variable coefficients.
We construct the corresponding wavelet representation.
As examples we consider wigglers and undulator magnets.
We consider the further modification of our variational
approach which may be applied in each scale.
\end{abstract}

\section{INTRODUCTION}
This is the third part of our eight presentations in which we consider
applications of methods from wavelet analysis to nonlinear accelerator
physics problems. This is a continuation of our results from [1]-[8],
which is based on our approach  to investigation
of nonlinear problems -- general, with additional structures (Hamiltonian,
symplectic or quasicomplex), chaotic, quasiclassical, quantum, which are
considered in the framework of local (nonlinear) Fourier analysis, or wavelet
analysis. Wavelet analysis is a relatively novel set of mathematical
methods, which gives us a possibility to work with well-localized bases in
functional spaces and with the general type of operators (differential,
integral, pseudodifferential) in such bases.
In this part we consider effects of insertion devices (section 2)
on beam dynamics.
In section 3 we consider generalization of our variational approach
 for the case of variable coefficients.
In section 4 we consider more  powerful variational approach
which is based on ideas of para-products and approximation
for multiresolution approach, which gives  us possibility
for computations in each scale separately.

\section{Effects of Insertion Devices on Beam Dynamics}
Assuming a sinusoidal field variation, we may consider according to [9]
the analytical treatment of the effects of insertion devices on beam dynamics.
One of the major detrimental aspects of the installation of insertion devices
is the resulting reduction of dynamic aperture. Introduction of non-linearities
leads to enhancement of  the amplitude-dependent tune shifts and distortion of
phase space. The nonlinear fields will produce significant effects at large
betatron amplitudes.
The components of the insertion device magnetic field used for the derivation of
equations of motion are as follows:
\begin{eqnarray}
B_x&=&\frac{k_x}{k_y}\cdot B_0 \sinh(k_xx)\sinh(k_yy)\cos(kz)\nonumber\\
B_y&=&B_0\cosh(k_xx)\cosh(k_yy)\cos(kz)\\
B_z&=&-\frac{k}{k_y}B_0\cosh(k_xx)\sinh(k_yy)\sin(kz),\nonumber
\end{eqnarray}
with $k_x^2+k_y^2=k^2=(2\pi/\lambda)^2$,
where $\lambda$ is  the period length of the insertion device, $B_0$ is
its magnetic field, $\rho$ is the radius of the curvature in the field $B_0$.
After a canonical transformation to change to
betatron variables, the Hamiltonian is averaged over the period of
the insertion device and hyperbolic functions are expanded to the fourth
order in $x$ and $y$ (or arbitrary order).
Then we have the following Hamiltonian:
\begin{eqnarray}
H&=&\frac{1}{2}[p_x^2+p_y^2]+\frac{1}{4k^2\rho^2}[k_x^2x^2+k_y^2y^2]\nonumber\\
 &+&\frac{1}{12k^2\rho^2}[k_x^4x^4+k_y^4y^4+3k_x^2k^2x^2y^2]\\
 &-&\frac{\sin(ks)}{2k\rho}[p_x(k_x^2x^2+k_y^2y^2)-2k_xp_yxy]\nonumber
\end{eqnarray}
We have in this case also nonlinear (polynomial with degree 3) dynamical system with variable
(periodic) coefficients. As related cases we may consider wiggler and undulator
magnets. We have in horizontal $x-s$ plane the following equations
\begin{eqnarray}
\ddot{x}&=&-\dot{s}\frac{e}{m\gamma}{B_z(s)}\\
\ddot{s}&=&\dot{x}\frac{e}{m\gamma}B_z(s),\nonumber
\end{eqnarray}
where magnetic field has periodic dependence on $s$ and hyperbolic on $z$.

\section{Variable Coefficients}
In the case when we have situation when our problem is described by a system of
nonlinear (polynomial)differential equations, we need to consider
extension of our previous approach which can take into  account
any type of variable coefficients (periodic, regular or singular).
We can produce such approach if we add in our construction additional
refinement equation, which encoded all information about variable
coefficients [10].
According to our variational approach we need to compute integrals of
the form
\begin{equation}\label{eq:var1}
\int_Db_{ij}(t)(\varphi_1)^{d_1}(2^m t-k_1)(\varphi_2)^{d_2}
(2^m t-k_2)\ud x,
\end{equation}
where now $b_{ij}(t)$ are arbitrary functions of time, where trial
functions $\varphi_1,\varphi_2$ satisfy a refinement equations:
\begin{equation}
\varphi_i(t)=\sum_{k\in{\bf Z}}a_{ik}\varphi_i(2t-k)
\end{equation}
If we consider all computations in the class of compactly supported wavelets
then only a finite number of coefficients do not vanish. To approximate
the non-constant coefficients, we need choose a different refinable function
$\varphi_3$ along with some local approximation scheme
\begin{equation}
(B_\ell f)(x):=\sum_{\alpha\in{\bf Z}}F_{\ell,k}(f)\varphi_3(2^\ell t-k),
\end{equation}
where $F_{\ell,k}$ are suitable functionals supported in a small neighborhood
of $2^{-\ell}k$ and then replace $b_{ij}$ in (\ref{eq:var1}) by
$B_\ell b_{ij}(t)$. In particular case one can take a characteristic function
and can thus approximate non-smooth coefficients locally. To guarantee
sufficient accuracy of the resulting approximation to (\ref{eq:var1})
it is important to have the flexibility of choosing $\varphi_3$ different
from $\varphi_1, \varphi_2$. In the case when D is some domain, we
can write
\begin{equation}
b_{ij}(t)\mid_D=\sum_{0\leq k\leq 2^\ell}b_{ij}(t)\chi_D(2^\ell t-k),
\end{equation}
where $\chi_D$ is characteristic function of D. So, if we take
$\varphi_4=\chi_D$, which is again a refinable function, then the problem of
computation of (\ref{eq:var1}) is reduced to the problem of calculation of
integral
\begin{eqnarray}
&&H(k_1,k_2,k_3,k_4)=H(k)=\nonumber\\
&&\int_{{\bf R}^s}\varphi_4(2^j t-k_1)\varphi_3(2^\ell t-k_2)\times\\
&&\varphi_1^{d_1}(2^r t-k_3)
\varphi_2^{d_2}(2^st-k_4)\ud x\nonumber
\end{eqnarray}
The key point is that these integrals also satisfy some sort of refinement
equation:
\begin{equation}
2^{-|\mu|}H(k)=\sum_{\ell\in{\bf Z}}b_{2k-\ell}H(\ell),\qquad \mu=d_1+d_2.
\end{equation}
This equation can be interpreted as the problem of computing an eigenvector.
Thus, we reduced the problem of extension of our method to the case of
variable coefficients to the same standard algebraical problem as in
the preceding sections.

So, the general scheme is the same one and we
have only one more additional
linear algebraic problem by which we in the same way can parameterize the
solutions of corresponding problem.
As example we demonstrate on Fig.~1 a simple model of (local) intersection
and the corresponding multiresolution representation (MRA).
\begin{figure}[ht]
\centering
\epsfig{file=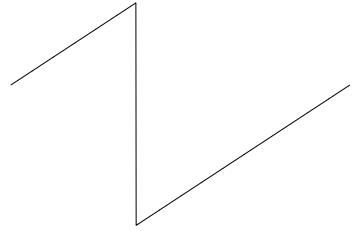,width=60.5mm, bb= 0 200 599 590, clip}
\caption{Simple insertion.}
\end{figure}
\begin{figure}[ht]
\centering
\epsfig{file=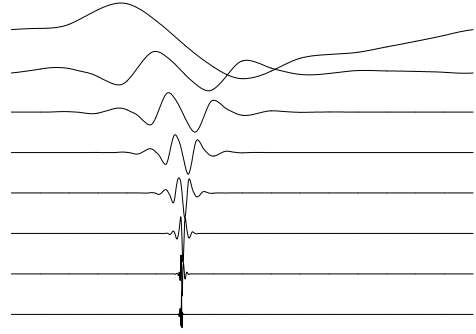, width=82.5mm, bb= 0 200 599 600, clip}
\caption{MRA representations.}
\end{figure}

\section{Evaluation of Nonlinearities Scale by Scale}
We consider  scheme of modification of our variational
approach in the case when we consider different scales separately. For this reason
we need to compute errors of approximations. The main problems come of course
from nonlinear terms. We follow the approach from [11].

Let $P_j$ be projection operators on the subspaces $V_j, j\in{\bf Z}$:
\begin{eqnarray}
P_j &:& L^2({\bf R}) \to V_j\\
(P_j f)(x)&=&\sum_k <f,\varphi_{j,k}> \varphi_{j,k}(x)\nonumber
\end{eqnarray}
and $Q_j$ are projection operators on the subspaces $W_j$:
\begin{eqnarray}
Q_j=P_{j-1}-P_j
\end{eqnarray}
So, for $u\in L^2({\bf R})$ we have $u_j=P_ju\quad$ and $u_j\in V_j$,
where $\{V_j\}, j\in{\bf Z}$ is a multiresolution analysis of $L^2({\bf R})$.
It is obviously that we can represent $u_0^2$ in the following form:
\begin{equation}\label{eq:form1}
u_0^2=2\sum^n_{j=1}(P_ju)(Q_ju)+\sum^n_{j=1}(Q_ju)(Q_ju)+u_n^2
\end{equation}
In this formula there is no interaction between different scales.
We may consider each term of (\ref{eq:form1}) as a bilinear mappings:
\begin{eqnarray}\label{eq:form2}
\displaystyle
M_{VW}^j : V_j\times W_j\to L^2({\bf R})=
V_j{\oplus_{j'\geq j}W_{j'}}
\end{eqnarray}
\begin{eqnarray}\label{eq:form3}
M_{WW}^j : W_j\times W_j\to L^2({\bf R})=V_j\oplus_{j'\geq j}W_{j'}
\end{eqnarray}
For numerical purposes we need formula (\ref{eq:form1}) with a finite number of
scales, but when we consider limits $j\to\infty$ we have
\begin{equation}
u^2=\sum_{j\in {\bf Z}}(2P_ju+Q_ju)(Q_ju),
\end{equation}
which is para-product of Bony, Coifman and Meyer.

Now we need to expand (\ref{eq:form1}) into the wavelet bases. To expand
each term in (\ref{eq:form1}) into wavelet basis, we need to consider
the integrals of the products of the basis functions, e.g.
\begin{equation}
M^{j,j'}_{WWW}(k,k',\ell)=\int^\infty_{-\infty}\psi^j_k(x)
\psi^j_{k'}(x)\psi^{j'}_\ell(x)\ud x,
\end{equation}
where $j'>j$ and
\begin{equation}
\psi^j_k(x)=2^{-j/2}\psi(2^{-j}x-k)
\end{equation}
are the basis functions.
If we consider compactly supported wavelets then
\begin{equation}
M_{WWW}^{j,j'}(k,k',\ell)\equiv 0\quad \mbox{for}\quad |k-k'|>k_0,
\end{equation}
where $k_0$ depends on the overlap of the supports of the basis functions
and
\begin{equation}\label{eq:form4}
|M_{WWW}^r(k-k',2^rk-\ell)|\leq C\cdot 2^{-r\lambda M}
\end{equation}
Let us define $j_0$ as the distance between scales such that for a given
$\varepsilon$ all the coefficients in (\ref{eq:form4}) with labels
$r=j-j'$, $r>j_0$ have absolute values less than $\varepsilon$. For the purposes of
computing with accuracy $\varepsilon$ we replace the mappings in
(\ref{eq:form2}), (\ref{eq:form3}) by
\begin{equation}\label{eq:z1}
M_{VW}^j : V_j\times W_j\to V_j\oplus_{j\leq j'\leq j_0}W_{j'}
\end{equation}
\begin{equation}\label{eq:z2}
M_{WW}^j : W_j\times W_j\to V_j\oplus_{J\leq j'\leq j_0}W_{j'}\nonumber
\end{equation}
Since
\begin{equation}
V_j\oplus_{j\leq j'\leq j_0}W_{j'}=V_{j_0-1}
\end{equation}
and
\begin{equation}
V_j\subset V_{j_0-1},\qquad W_j\subset V_{j_0-1}
\end{equation}
we may consider bilinear mappings (\ref{eq:z1}), (\ref{eq:z2}) on
$V_{j_0-1}\times V_{j_0-1}$.
For the evaluation of (\ref{eq:z1}), (\ref{eq:z2}) as mappings
$V_{j_0-1}\times V_{j_0-1}\to V_{j_0-1}$
we need significantly fewer coefficients than for
mappings (\ref{eq:z1}), (\ref{eq:z2}). It is enough to consider only
coefficients
\begin{equation}
M(k,k',\ell)=2^{-j/2}\int^\infty_\infty\varphi(x-k)\varphi(x-k')\varphi(x-\ell)\ud
x,
\end{equation}
where $\varphi(x)$ is scale function. Also we have
\begin{equation}
M(k,k',\ell)=2^{-j/2}M_0(k-\ell,k'-\ell),
\end{equation}
where
\begin{equation}
M_0(p,q)=\int\varphi(x-p)\varphi(x-q)\varphi(x)\ud x
\end{equation}
Now as in section (3) we may derive and solve a system of linear equations to
find $M_0(p,q)$ and obtain explicit representation for solution.

We are very grateful to  M.~Cornacchia (SLAC),
W.~He\-rrmannsfeldt (SLAC),
 Mrs. M.~Laraneta (UCLA), J.~Ko-\\
no (LBL)
for their permanent encouragement.

 \end{document}